\documentclass[aps,twocolumn,showpacs,preprintnumbers,floats]{revtex4}\def\@cite#1#2{\textsuperscript{[{#1\if@tempswa , #2\fi}]}}
\usepackage{mathrsfs}
\usepackage{longtable,lscape}
\usepackage{txfonts}
\usepackage{amssymb}
\usepackage{indentfirst}
\usepackage{graphicx,booktabs}
\usepackage{float}
\usepackage{setspace}
\usepackage{epstopdf}

\usepackage{multirow}
\usepackage{color}
\usepackage{epsfig}

\begin{document}


\title{A possible explanation of the threshold enhancement in the process $e^+e^-\rightarrow \Lambda\bar{\Lambda}$}

\author{Li-Ye Xiao$^{1,2}$~\footnote {E-mail: lyxiao@pku.edu.cn}, Xin-Zhen Weng$^{1}$~\footnote {E-mail: xzhweng@pku.edu.cn}, Xian-Hui Zhong$^{3,4}$~\footnote {E-mail: zhongxh@hunnu.edu.cn} and Shi-Lin Zhu$^{1,2,5}$~\footnote {E-mail: zhusl@pku.edu.cn}}

\affiliation{ 1) School of Physics and State Key Laboratory of
Nuclear Physics and Technology, Peking University, Beijing 100871,
China } \affiliation{ 2)  Center of High Energy Physics, Peking
University, Beijing 100871, China} \affiliation{ 3) Department of
Physics, Hunan Normal University, and Key Laboratory of
Low-Dimensional Quantum Structures and Quantum Control of Ministry
of Education, Changsha 410081, China } \affiliation{ 4) Synergetic
Innovation Center for Quantum Effects and Applications (SICQEA),
Hunan Normal University, Changsha 410081, China} \affiliation{ 5)
Collaborative Innovation Center of Quantum Matter, Beijing 100871,
China}


\begin{abstract}
Inspired by the recent measurement of the process $e^+e^-\rightarrow
\Lambda\bar{\Lambda}$, we calculate the mass spectrum of the $\phi$
meson with the GI model. For the excited vector strangeonium states
$\phi(3S,~4S,~5S,~6S)$ and $\phi(2D,~3D,~4D,~5D)$, we further
investigate the electronic decay width with the Van Royen-Weisskopf
formula, and the partial widths of the $\Lambda\bar{\Lambda}$,
$\Xi^{-(*)}\bar{\Xi}^+$, and $\Sigma^{+(*)}\bar{\Sigma}^{-(*)}$
decay modes with the extended quark pair creation model. We find
that the electronic decay width of the $D$-wave vector strangeonium
is about $3\sim8$ times larger than that of the $S$-wave vector
strangeonium. Around 2232 MeV the partial decay width of the
$\Lambda\bar{\Lambda}$ mode can reach up to several MeV for
$\phi(3^3S_1)$, while the partial $\Lambda\bar{\Lambda}$ decay width
of $\phi(2^3D_1)$ is $\mathcal{O}(10^{-3})$ keV. If the threshold
enhancement reported by the BESIII Collaboration arises from the
strangeonium meson, this state is very likely to be the
$\phi(3^3S_1)$ state. We also note that the $\Lambda\bar{\Lambda}$
and $\Sigma^{+}\bar{\Sigma}^{-}$ partial decay widths of the states
$\phi(3^3D_1)$ and $\phi(4^3S_1)$ are about several MeV,
respectively, which are enough to be observed in future experiments.

\end{abstract}


\maketitle

\section{Introduction}

Because the timelike electromagnetic form factors (FFs) provide a
key to understand the strong interactions and inner structure of
hadrons, there have been many measurements via the process
$e^+e^-\rightarrow B\bar{B}$
~\cite{Delcourt:1979ed,Antonelli:1998fv,Armstrong:1992wq,Pedlar:2005sj,Ablikim:2017lct}
(where $B$ stands for a spin-1/2 ground baryon state). The
non-vanishing cross section in the near-threshold region has been
observed
~\cite{Lees:2013ebn,Bardin:1994am,Achasov:2014ncd,Aubert:2007uf}.
Unusual behavior near threshold implies a more complicated
underlying physical scenario and has driven many theoretical
interpretations, including $B\bar{B}$ bound states or meson-like
resonances~\cite{ElBennich:2008vk,Haidenbauer:2006dm,Zhao:2013ffn,Haidenbauer:2016won,Fonvieille:2009px,Kang:2015yka,Cao:2018kos,Yang:2019mzq},
final-state interactions~\cite{Haidenbauer:2014kja,Dalkarov:2009yf}
and an attractive Coulomb
interaction~\cite{Baldini:2007qg,Ferroli:2010bi}.

Very recently, the BESIII Collaboration studied the process
$e^+e^-\rightarrow \Lambda\bar{\Lambda}$ with improved
precision~\cite{Ablikim:2017pyl}. The Born cross section at
$\sqrt{s}=2.2324$ GeV, which is 1.0 MeV above the
$\Lambda\bar{\Lambda}$ mass threshold, is measured to be
$305\pm45^{+66}_{-36}$ pb. Is the unexpected feature in the
near-threshold region due to an unobserved strangeonium meson
resonance? In the present work, we will try to answer this question.

We will calculate the spectrum of the $s\bar{s}$ system in the
framework of the Godfrey-Isgur (GI) model~\cite{Godfrey:1985xj},
which has achieved a good description of the known mesons and
baryons~\cite{Godfrey:1985xj,Godfrey:2004ya,Capstick:1986bm}. After
we obtain the masses of the higher excited strangeonium states, we
further estimate the electronic decay width of the $J^{PC}=1^{--}$
states $\phi(2D,~3D,~4D,~5D)$ and $\phi(3S,~4S,~5S,~6S)$ with the
Van Royen-Weisskopf formula~\cite{VanRoyen:1967nq}. Meanwhile, we
use the extended quark pair creation
model~\cite{Xiao:2018iez,Weng:2018ebv} to calculate the partial
$\Lambda\bar{\Lambda}$, $\Xi^{-(*)}\bar{\Xi}^+$, and
$\Sigma^{+(*)}\bar{\Sigma}^{-(*)}$ decay widths of those vector
states with the obtained spatial wave functions. Considering there
existing many theoretical calculations of the two-body strong decays
of the $s\bar{s}$ system with various models in the
literature~\cite{Pang:2019ttv,Ebert:2014jxa,Ding:2007pc,Barnes:1996ff,Barnes:2002mu,Ricken:2003ua},
in the present work we will emphasise on the baryon-antibaryon decay
mode and electronic decay properties.

According to the theoretical predictions from various models, the
masses of $\phi(3^3S_1)$ and $\phi(2^3D_1)$ mesons are about 2.2 GeV
(see Table~\ref{MASS}). Therefore, we calculate the $e^+e^-$ and
$\Lambda\bar{\Lambda}$ partial decay widths of the excited vector
states $\phi(3^3S_1)$ and $\phi(2^3D_1)$. We find that the
electronic decay width of $\phi(3^3S_1)$ is about 1/3 times smaller
than that of $\phi(2^3D_1)$. However around 2232 MeV the partial
decay width of the $\Lambda\bar{\Lambda}$ mode can reach up to
several MeV for $\phi(3^3S_1)$, while the partial
$\Lambda\bar{\Lambda}$ decay width of the states $\phi(2^3D_1)$ is a
very small value $\mathcal{O}(10^{-3})$ keV. The threshold
enhancement in the process $e^+e^-\rightarrow \Lambda\bar{\Lambda}$
observed by the BESIII Collaboration~\cite{Ablikim:2017pyl} may be
caused by $\phi(3^3S_1)$. We also notice that the
$\Lambda\bar{\Lambda}$ and $\Sigma^{+}\bar{\Sigma}^{-}$ partial
decay widths of the states $\phi(3^3D_1)$ and $\phi(4^3S_1)$ are
about several MeV, respectively. These two states have a good
potential to be observed in future experiments via their
corresponding main baryon-antibaryon decay channel.

This paper is organized as follows. In Sec. II we give a brief
introduction of the GI model and calculate the spectrum of the
$s\bar{s}$ system. Then we present the Van Royen-Weisskopf formula
and give the electronic decay properties in Sec. III.  In Sec. IV we
discuss the extended quark pair creation model and baryon-antibaryon
decay results. We give a short summary in Sec. V.

\begin{figure*}[]
\centering \epsfxsize=14 cm \epsfbox{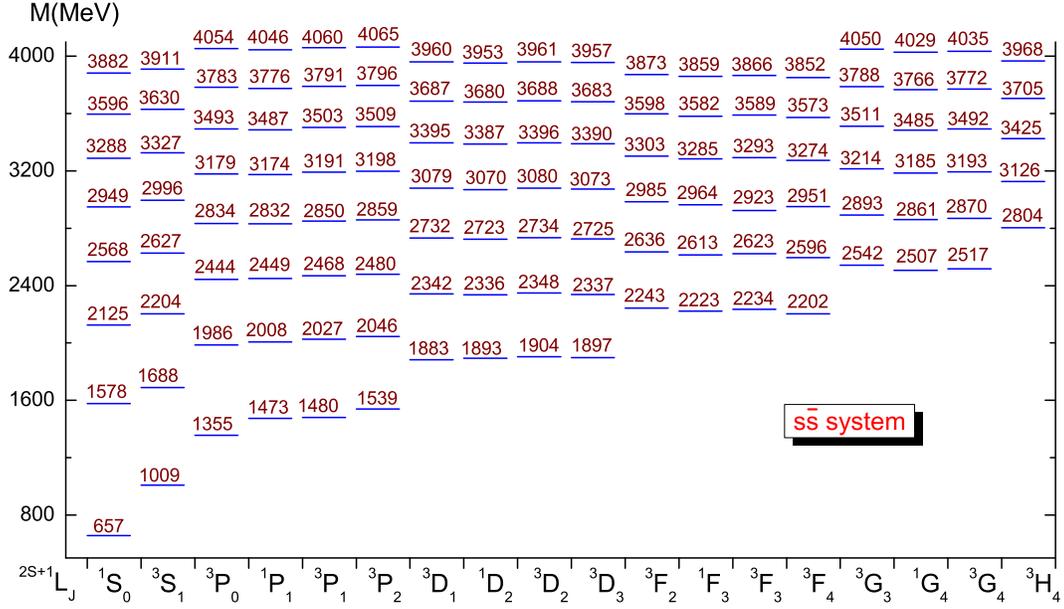} \caption{The spectrum
of the $s\bar{s}$ system.}\label{MASS1}
\end{figure*}

\begin{table*}[ht]
\caption{\label{MASS} The predicted masses (MeV) of the higher
$\phi$ mesons with $J^{PC}=1^{--}$ from various models. }
\begin{tabular}{ccccccccccccccc}\hline\hline
~~~~~~~~&\text{This work}~~~~~~~~&~~~~~~~~&~~~~~~~~&~~~~~~~~&~~~~~~~~&\\
\text{State}~~~~~~~~&\text{Mass}~~~~~~~~&\text{MGI}~\cite{Pang:2019ttv}~~~~~~~~&\text{GI}~\cite{Godfrey:1985xj}~~~~~~~~&\text{RQM}~\cite{Ebert:2009ub}~~~~~~~~&\text{COQM}~\cite{Ishida:1986vn}
~~~~~~~~&\text{Exp.}~\cite{Tanabashi:2018oca}\\
\hline
$\phi(1^3S_1)$~~~~~~~~&1009~~~~~~~~&1030~~~~~~~~&1020~~~~~~~~&1038~~~~~~~~&1020~~~~~~~~&1020\\
$\phi(2^3S_1)$~~~~~~~~&1688~~~~~~~~&1687~~~~~~~~&1690~~~~~~~~&1698~~~~~~~~&1740~~~~~~~~&1680\\
$\phi(3^3S_1)$~~~~~~~~&2204~~~~~~~~&2149~~~~~~~~&$\cdot\cdot\cdot$~~~~~~~~&2119~~~~~~~~&2250~~~~~~~~&--\\
$\phi(4^3S_1)$~~~~~~~~&2627~~~~~~~~&2498~~~~~~~~&$\cdot\cdot\cdot$~~~~~~~~&2472~~~~~~~~&2540~~~~~~~~&--\\
$\phi(5^3S_1)$~~~~~~~~&2996~~~~~~~~&$\cdot\cdot\cdot$~~~~~~~~&$\cdot\cdot\cdot$~~~~~~~~&2782~~~~~~~~&$\cdot\cdot\cdot$~~~~~~~~&--\\
$\phi(6^3S_1)$~~~~~~~~&3327~~~~~~~~&$\cdot\cdot\cdot$~~~~~~~~&$\cdot\cdot\cdot$~~~~~~~~&$\cdot\cdot\cdot$~~~~~~~~&$\cdot\cdot\cdot$~~~~~~~~&--\\
$\phi(1^3D_1)$~~~~~~~~&1883~~~~~~~~&1869~~~~~~~~&$\cdot\cdot\cdot$~~~~~~~~&1845~~~~~~~~&1750~~~~~~~~&--\\
$\phi(2^3D_1)$~~~~~~~~&2342~~~~~~~~&2276~~~~~~~~&$\cdot\cdot\cdot$~~~~~~~~&2258~~~~~~~~&2260~~~~~~~~&--\\
$\phi(3^3D_1)$~~~~~~~~&2732~~~~~~~~&2593~~~~~~~~&$\cdot\cdot\cdot$~~~~~~~~&2607~~~~~~~~&$\cdot\cdot\cdot$~~~~~~~~&--\\
$\phi(4^3D_1)$~~~~~~~~&3079~~~~~~~~&$\cdot\cdot\cdot$~~~~~~~~&$\cdot\cdot\cdot$~~~~~~~~&$\cdot\cdot\cdot$~~~~~~~~&$\cdot\cdot\cdot$~~~~~~~~&--\\
$\phi(5^3D_1)$~~~~~~~~&3395~~~~~~~~&$\cdot\cdot\cdot$~~~~~~~~&$\cdot\cdot\cdot$~~~~~~~~&$\cdot\cdot\cdot$~~~~~~~~&$\cdot\cdot\cdot$~~~~~~~~&--\\
\hline\hline
\end{tabular}
\end{table*}

\section{mass spectrum}

In this work, we employ the GI model to calculate the mass spectrum
of the higher excited strangeonium. According to the GI
model~\cite{Godfrey:1985xj}, the Hamiltonian between the quark and
antiquark reads
\begin{equation}
\tilde{H}= \left(\mathbf{p}^2+m_1^2\right)^{1/2}
+\left(\mathbf{p}^2+m_2^2\right)^{1/2}
+V_{\text{eff}}(\mathbf{p},\mathbf{r}),
\end{equation}
where $m_i$ and $\mathbf{p}$ are the quark's mass and momentum in
the center-of-mass frame. $V_{\text{eff}}(\mathbf{p},\mathbf{r})$ is
the potential between the quark and antiquark, which can be obtained
by the on-shell scattering amplitude between the quark and antiquark
in the center-of-mass frame. This Hamiltonian contains the
short-range $\gamma^{\mu}\otimes\gamma_{\mu}$ one-gluon-exchange
(OGE) interaction and the $1\otimes1$ linear confining interaction
suggested by lattice QCD. In the nonrelativistic limit, it can
reduce to the familiar Breit-Fermi interaction
\begin{equation}
V_{\text{eff}}(\mathbf{p},\mathbf{r})={H}_{12}^{\text{conf}}+{H}_{12}^{\text{hyp}}+{H}_{12}^{\text{so}}.
\end{equation}
Here, ${H}_{12}^{\text{conf}}$ is the spin-independent linear
confinement and Coulomb-type interaction; ${H}_{12}^{\text{hyp}}$ is
the color-hyperfine interaction and ${H}_{12}^{\text{so}}$ is the
spin-orbit interaction.

To incorporate the relativistic effects, Godfrey and Isgur further
built a semiquantitative model~\cite{Godfrey:1985xj}. By introducing
the smearing function for a meson $q_i\bar{q}_j$
\begin{equation}
\rho_{ij}(\mathbf{r}-\mathbf{r}') =
\frac{\sigma_{ij}^3}{\pi^{3/2}}\mathrm{e}^{-\sigma_{ij}^2(\mathbf{r}-\mathbf{r}')^2},
\end{equation}
the OGE potential $G(r)=-4\alpha(r)/3r$ and confining potential
$S(r)=br+c$ are smeared to $\tilde{G}(r)$ and $\tilde{S}(r)$ via
\begin{equation}
\tilde{f}(r) =
\int\mathrm{d}^3\mathbf{r}'\rho_{ij}(\mathbf{r}-\mathbf{r}')f(r').
\end{equation}
Through the introduction of the momentum-dependent factors, the
Coulomb term is modified according to
\begin{equation}
\tilde{G}(r) {\to} \left(1+\frac{p^2}{E_1E_2}\right)^{1/2}
\tilde{G}(r) \left(1+\frac{p^2}{E_1E_2}\right)^{1/2},
\end{equation}
and the contact, tensor, vector spin-orbit, and scalar spin-orbit
potentials were modified according to
\begin{equation}
\frac{\tilde{V}_{i}(r)}{m_1m_2} {\to}
\left(\frac{m_1m_2}{E_1E_2}\right)^{1/2+\epsilon_i}
\frac{\tilde{V}_{i}(r)}{m_1m_2}
\left(\frac{m_1m_2}{E_1E_2}\right)^{1/2+\epsilon_i}
\end{equation}
where $\epsilon_{i}$ corresponds to the  contact (c), tensor (t),
vector spin-orbit [so(v)], and scalar spin-orbit [so(s)].

With the notation
\begin{equation}
f_{\alpha\beta}^{i}(r) =
\left(\frac{m_{\alpha}m_{\beta}}{E_{\alpha}E_{\beta}}\right)^{1/2+\epsilon_{i}}
f(r)
\left(\frac{m_{\alpha}m_{\beta}}{E_{\alpha}E_{\beta}}\right)^{1/2+\epsilon_{i}}
\end{equation}
we have
\begin{equation}
V_{\text{eff}}(\mathbf{p},\mathbf{r})=\tilde{H}_{12}^{\text{conf}}+\tilde{H}_{12}^{\text{hyp}}+\tilde{H}_{12}^{\text{so}},
\end{equation}
where
\begin{widetext}
\begin{equation}
\tilde{H}_{12}^{\text{conf}}=
\left(1+\frac{p^2}{E_1E_2}\right)^{1/2} \tilde{G}(r)
\left(1+\frac{p^2}{E_1E_2}\right)^{1/2} +\tilde{S}(r),
\end{equation}
\begin{equation}
\tilde{H}_{12}^{\text{so}}=
\frac{\mathbf{S}_1\cdot\mathbf{L}}{2m_1^2}\frac{1}{r}\frac{\partial\tilde{G}_{11}^{\text{so}(v)}}{\partial{r}}
+\frac{\mathbf{S}_2\cdot\mathbf{L}}{2m_2^2}\frac{1}{r}\frac{\partial\tilde{G}_{22}^{\text{so}(v)}}{\partial{r}}
+\frac{\left(\mathbf{S}_1+\mathbf{S}_2\right)\cdot\mathbf{L}}{m_1m_2}\frac{1}{r}\frac{\partial\tilde{G}_{12}^{\text{so}(v)}}{\partial{r}}
-\frac{\mathbf{S}_1\cdot\mathbf{L}}{2m_1^2}\frac{1}{r}\frac{\partial\tilde{S}_{11}^{\text{so}(s)}}{\partial{r}}
-\frac{\mathbf{S}_2\cdot\mathbf{L}}{2m_2^2}\frac{1}{r}\frac{\partial\tilde{S}_{22}^{\text{so}(s)}}{\partial{r}},
\end{equation}
and
\begin{equation}
\tilde{H}_{12}^{\text{hyp}}=
\frac{2\mathbf{S}_{i}\cdot\mathbf{S}_{j}}{3m_1m_2}\nabla^2\tilde{G}_{12}^{\text{c}}
-\left(\frac{\mathbf{S}_{1}\cdot{\hat{r}}\mathbf{S}_{2}\cdot\hat{r}-\frac{1}{3}\mathbf{S}_{1}\cdot\mathbf{S}_{2}}{m_1m_2}\right)
\left(\frac{\partial^2}{\partial{r}^2}-\frac{1}{r}\frac{\partial}{\partial{r}}\right)\tilde{G}_{12}^{\text{t}}.
\end{equation}
\end{widetext}
The spin-orbit term $\tilde{H}_{12}^{\text{so}}$ can be decomposed
into a symmetric part $\tilde{H}_{(12)}^{\text{so}}$ and an
antisymmetric part $\tilde{H}_{[12]}^{\text{so}}$, while the
$\tilde{H}_{[12]}^{\text{so}}$ vanishes when $m_1=m_2$.

We adopt the free parameters in the original work of the GI
model~\cite{Godfrey:1985xj}, and diagonalize the Hamiltonian in the
simple harmonic oscillator bases $|n ^{2S+1}L_{J}\rangle$. The
resulting mass spectrum of the strangeonium are shown in
Fig.~\ref{MASS1}. Meanwhile, we compare our predicted mass of the
higher vector $\phi$ mesons with various models predictions, as
listed in Table~\ref{MASS}.

\section{The electronic decays}
With the Van Royen-Weisskopf
formula~\cite{VanRoyen:1967nq,Li:2009zu}, the electronic decay width
of the excited vector strangeonium states  is given by
\begin{eqnarray}\label{EE1}
\Gamma[\phi(nS)\rightarrow e^+e^-]\propto \frac{4\alpha^2e^2_s}{M^2_{nS}}|R_{nS}(0)|^2,\\
\Gamma[\phi(nD)\rightarrow e^+e^-]\propto
\frac{25\alpha^2e^2_s}{2M^2_{nD}m^4_s}|R{''}_{nD}(0)|^2.
\end{eqnarray}
Here, $\alpha=\frac{1}{137}$ denotes the fine structure constant.
$m_s=450$ MeV and $e_s=-\frac{1}{3}$ are the strange quark
constituent mass and charge in unit of electron charge,
respectively. $M_{nS}(M_{nD})$ is the mass for $\phi(nS)(\phi(nD))$.
$R_{nS}(0)$ represents the radial $S$ wave function at the origin,
and $R{''}_{nD}(0)$ represents the second derivative of the radial
$D$ wave function at the origin.

In the present calculation, we adopt the simple harmonic oscillator
(SHO) wave functions for the space-wave functions of the initial
meson. According to the wave functions obtained in mass spectrum
calculations, we get the root mean square radius of the vector
states. Then, we determine the value of harmonic oscillator strength
$\beta_{th}$ between the two strange quarks for the initial mesons
(as listed in Table~\ref{EE3}).

According to PDG~\cite{Tanabashi:2018oca}, the electronic decay
branching ratio for $\phi(1S)$ is
\begin{eqnarray}\label{EE2}
\mathcal{B}[\phi(1S)\rightarrow
e^+e^-]=(2.973\pm0.034)\times10^{-4}.
\end{eqnarray}
Combining this ratio with its total decay
widths($\Gamma=4.249\pm0.013$ MeV), the central value of the
electronic decay width is $\Gamma[\phi(1S)\rightarrow e^+e^-]=1.26$
keV. Then, from the formulas~(12)-(13), we can obtain electronic
decay width ratios of between the higher excited vector strangeonium
states and the state $\phi(1S)$. Thus, we can get those states
electronic decay widths, as shown in Table~\ref{EE3}.

\begin{table}[ht]
\caption{\label{EE3} The predicted electronic decay widths of the
higher $\phi$ mesons with $J^{PC}=1^{--}$. The unit is MeV for the
mass and harmonic oscillator strength $\beta_{th}$. The unit of the
$e^+e^-$ decay width is keV. $R$ is the electronic decay width ratio
between the higher excited states and the state $\phi(1S)$.
$\Gamma[e^+e^-]=R\times\Gamma[\phi(1S)\rightarrow e^+e^-]$ denotes
the electronic decay width for each state.}
\begin{tabular}{ccccccccccccccc}\hline\hline
\text{State}~~~~~~~~&\text{Mass}~~~~~~~~&$\beta_{th}$~~~~~~~~&$R$~~~~~~~~&$\Gamma[e^+e^-]$(keV)\\
\hline
$\phi(3^3S_1)$~~~~~~~~&2204~~~~~~~~&368~~~~~~~~&0.17~~~~~~~~&0.21\\
$\phi(4^3S_1)$~~~~~~~~&2627~~~~~~~~&351~~~~~~~~&0.12~~~~~~~~&0.15\\
$\phi(5^3S_1)$~~~~~~~~&2996~~~~~~~~&341~~~~~~~~&0.09~~~~~~~~&0.11\\
$\phi(6^3S_1)$~~~~~~~~&3327~~~~~~~~&334~~~~~~~~&0.08~~~~~~~~&0.10\\
$\phi(2^3D_1)$~~~~~~~~&2342~~~~~~~~&375~~~~~~~~&0.47~~~~~~~~&0.59\\
$\phi(3^3D_1)$~~~~~~~~&2732~~~~~~~~&355~~~~~~~~&0.54~~~~~~~~&0.68\\
$\phi(4^3D_1)$~~~~~~~~&3079~~~~~~~~&344~~~~~~~~&0.61~~~~~~~~&0.77\\
$\phi(5^3D_1)$~~~~~~~~&3395~~~~~~~~&336~~~~~~~~&0.70~~~~~~~~&0.88\\
\hline\hline
\end{tabular}
\end{table}

\begin{figure*}[hpbt]
\centering \epsfxsize=16 cm \epsfbox{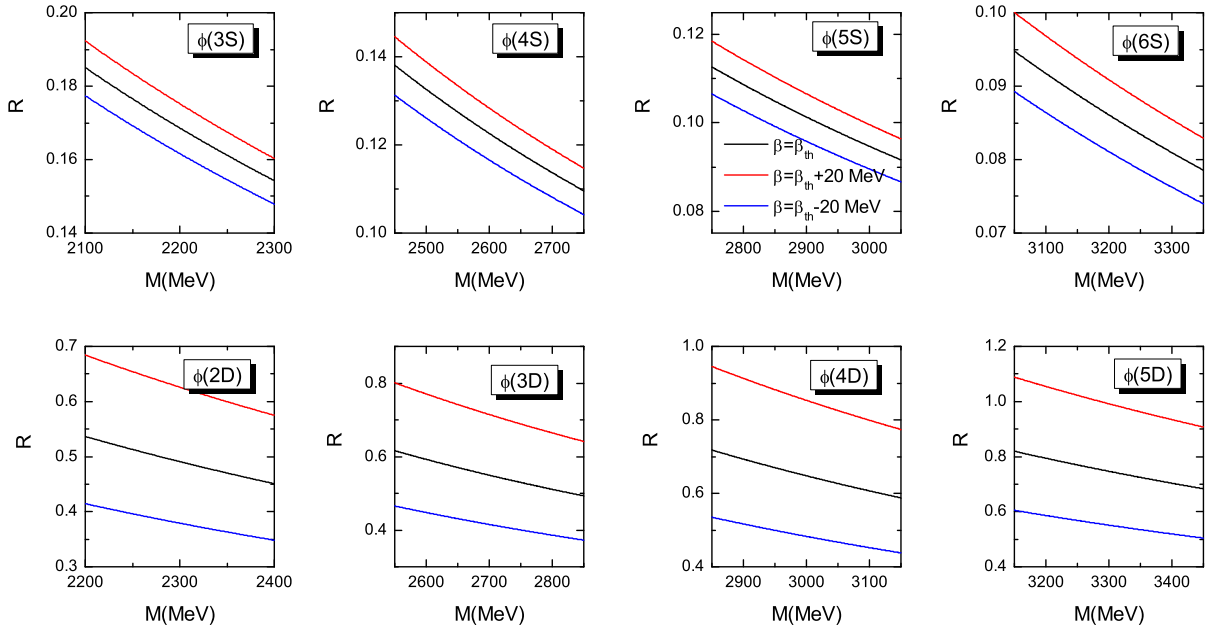} \caption{The
variation of the electronic decay width ratio with the mass of the
S- and D-wave vector strangeonium. The red, black, and blue lines
correspond to the predictions with different values of
$\beta$=$\beta_{th}+20$ MeV, $\beta_{th}$, and $\beta_{th}-20$ MeV,
respectively. }\label{SDWAVEEE}
\end{figure*}

From the table, the ratio $R$ is smaller than one. The electronic
decay widths of the excited vector strangeonium states
$\phi(3S,~4S,~5S,~6S)$ and $\phi(2D,~3D,~4D,~5D)$ are smaller than
that of the state $\phi(1S)$. Meanwhile, the electronic decay width
of the $D$-wave vector strangeonium is about $3\sim8$ times larger
than that of the $S$-wave vector strangeonium. For the $S$-wave
states, our predictions are in accordance with
ref.~\cite{Badalian:2019xir}, while for the $D$-wave states, our
predictions are about 3 times larger than those of
ref.~\cite{Badalian:2019xir}.

Considering the uncertainties of the predicted mass and harmonic
oscillator strength $\beta_{th}$, we plot the variation of the
electronic decay width ratio $R$ as a function of the mass with
different values of $\beta$=$\beta_{th}+20$ MeV, $\beta_{th}$, and
$\beta_{th}-20$ MeV, respectively, in Fig.~\ref{SDWAVEEE}. It is
obvious that the ratio $R$ decreases with the mass with the same
$\beta$ values.

\section{Double baryon decay mode}\label{three}
\subsection{The $^3P_0$ model}
The quark pair creation ($^3P_0$) model was first proposed by
Micu~\cite{Micu:1968mk}, Carlitz and
Kislinger~\cite{Carlitz:1970xb}, and further developed by the Orsay
group~\cite{LeYaouanc:1972vsx,LeYaouanc:1988fx,LeYaouanc:1977fsz},
which has been widely used to study the OZI-allowed two-body strong
decays of hadrons. Very recently, the $^3P_0$ model was extended to
study some OZI-allowed three-body strong decays~\cite{Weng:2018ebv}
as well. In the framework of this  model, the interaction
Hamiltonian for one quark pair creation was described
as~\cite{Geiger:1994kr,Ackleh:1996yt,Close:2005se}
\begin{eqnarray}
H_{q\bar{q}}=\gamma\sum_f2m_f\int d^3x\bar{\psi}_f\psi_f.
\end{eqnarray}
Here, $\gamma$ is a dimensionless parameter and usually determined
by fitting the experimental data. $m_f$ denotes the constituent
quark mass of flavor $f$  and $\psi_f$ stands for a Dirac quark
field.

\begin{figure}[htpb]
\centering \epsfxsize=5.8 cm \epsfbox{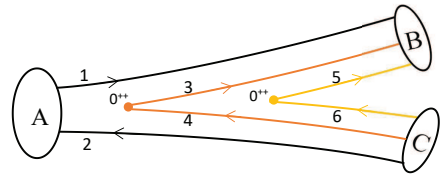} \caption{The
strangeonium system decays into double baryons.}\label{qkp}
\end{figure}

In our previous work~\cite{Xiao:2018iez}, we extended the $^3P_0$
model to study the partial decay width of the
$\Lambda_c\bar{\Lambda}_c$ mode for the charmonium system. In this
work, we further use this model to study the process
$\phi^*(A)\rightarrow B(B)+\bar{B}(C)$, where $\phi^*$ denotes the
excited strangeonium states. As pointed out in
Ref.~\cite{Xiao:2018iez}, two light quark pairs should be created
for this type of reaction (as shown in Fig.~\ref{qkp}), and the
helicity amplitude $M^{M_{J_A}M_{J_B}M_{J_B}}$ reads
\begin{eqnarray}\label{Q1}
&\delta^3(\mathbf{p}_A-\mathbf{p}_B-\mathbf{p}_C)M^{M_{J_A}M_{J_B}M_{J_B}} \nonumber\\
&=\sum_k\frac{\langle BC|H_{q\bar{q}}|k\rangle\langle
k|H_{q\bar{q}}|A\rangle}{E_k-E_A}.
\end{eqnarray}
Here, $\mathbf{p}_I(I=A,~B,~C)$ denotes the momentum of the hadron.
$E_{A(k)}$ represents the energy of the initial(intermediate) state
$A(k)$. Considering the quark-hadron duality~\cite{Shifman:2000jv},
we simplify the calculations via taking $E_k-E_A$ as a constant,
namely $E_k-E_A\approx2m_{q}$. Here, $m_q$ is the constituent quark
mass of the created quark. We adopt this crude approximation because
the intermediate state differs from the initial state by two created
additional quarks at the quark
level~\cite{Xiao:2018iez,Weng:2018ebv}. Thus, we can rewrite the
Eq.~(\ref{Q1}) as
\begin{eqnarray}\label{Q2}
\delta^3(\mathbf{p}_A-\mathbf{p}_B-\mathbf{p}_C)M^{M_{J_A}M_{J_B}M_{J_B}}\approx\frac{\langle
BC|H_{q\bar{q}}H_{q\bar{q}}|A\rangle}{2m_{q}}.
\end{eqnarray}
Then, the transition operator for the two quark pairs creation in
the nonrelativistic limit reads
\begin{eqnarray}
T&=&\frac{9\gamma^2}{2m_{q}} \sum_{m,m'}\langle1m;1-m|00\rangle\langle1m';1-m'|00\rangle\\
\nonumber &&\int
d^3\mathbf{p}_3d^3\mathbf{p}_4d^3\mathbf{p}_5d^3\mathbf{p}_6\delta^3(\mathbf{p}_3+\mathbf{p}_4)\delta^3(\mathbf{p}_5+\mathbf{p}_6)\\
\nonumber &&\times\varphi^{34}_0\omega^{34}_0\chi^{34}_{1,-m}
\mathcal{Y}_1^m(\frac{\mathbf{p}_3-\mathbf{p}_4}{2})a^{\dagger}_{3i}b^{\dagger}_{4j}\\
\nonumber &&\times\varphi^{56}_0\omega^{56}_0\chi^{56}_{1,-m'}
\mathcal{Y}_1^{m'}(\frac{\mathbf{p}_5-\mathbf{p}_6}{2})a^{\dagger}_{5i}b^{\dagger}_{6j},
\end{eqnarray}
where $\mathbf{p}_i$ ($i$=3,~4,~5,~6) stands for the three-vector
momentum of the $i$th quark.
$\varphi_0=(u\bar{u}+d\bar{d}+s\bar{s})/\sqrt{3}$ corresponds to the
flavor function and $\omega_0=\delta_{ij}$ represents the color
singlet of the quark pairs created from vacuum. $\chi_{1,-m(m')}$
are the spin triplet states for the created quark pairs. The solid
harmonic polynomial
$\mathcal{Y}_1^{m(m')}(\mathbf{p})\equiv|\mathbf{p}|Y^{m(m')}_1(\theta_p,\phi_p)$
denotes the $P$-wave quark pairs. $a^{\dagger}_{i}b^{\dagger}_{j}$
is the creation operator representing the quark pair creation in the
vacuum.

Finally, the hadronic decay width $\Gamma$ in the relativistic phase
space reads
\begin{eqnarray}
\Gamma[A\rightarrow
BC]=\pi^2\frac{|\mathbf{p}|}{M_A^2}\frac{1}{2J_A+1}\sum_{M_{J_A},M_{J_B},M_{J_C}}|\mathcal{M}^{M_{J_A}M_{J_B}M_{J_C}}|^2.
\end{eqnarray}
Here, $\mathbf{p}$ represents the momentum of the daughter baryon.
$M_A$ and $J_A$ are the mass and total angular quantum number of the
parent baryon $A$, respectively. In the center of mass frame of the
parent baryon $A$, $\mathbf{p}$ reads
\begin{eqnarray}
|\mathbf{p}|=\frac{\sqrt{[M_A^2-(M_B-M_C)^2][M_A^2-(M_B+M_C)^2]}}{2M_A}.
\end{eqnarray}

\begin{table}[h]
\caption{\label{paraments} The parameters we used in this
work~\cite{Tanabashi:2018oca,Godfrey:2015dva}. The unit is MeV
except the $\gamma$, which has no unit. }
\begin{tabular}{lccllll}\hline\hline
Mass of the final baryon ~~~~~~~~&$\Lambda$~~~~~~~~&1115.68\\
                          ~~~~~~~~&$\bar{\Lambda}$~~~~~~~~&1115.68\\
                          ~~~~~~~~&$\Sigma^+$~~~~~~~~&1189.37\\
                          ~~~~~~~~&$\bar{\Sigma}^-$~~~~~~~~&1189.37\\
                          ~~~~~~~~&$\Xi^-$~~~~~~~~&1321.71\\
                          ~~~~~~~~&$\bar{\Xi}^+$~~~~~~~~&1321.71\\
                          ~~~~~~~~&$\Sigma^{+*}$~~~~~~~~&1382.80\\
                          ~~~~~~~~&$\bar{\Sigma}^{-*}$~~~~~~~~&1382.80\\
                          ~~~~~~~~&$\Xi^{-*}$~~~~~~~~&1535.0\\
\hline
Constituent quark mass  ~~~~~~~~&$m_u$~~~~~~~~&330\\
                        ~~~~~~~~&$m_d$~~~~~~~~&330\\
                        ~~~~~~~~&$m_s$~~~~~~~~&450\\
\hline
Harmonic oscillator parameter~~~~~~~~&$\alpha$~~~~~~~~&400\\
\hline
Strength of the quark pair ~~~~~~~~&$\gamma$~~~~~~~~&6.95\\
creation from the vacuum   ~~~~~~~~&~~~~~~~~&\\
\hline\hline
\end{tabular}
\end{table}

In our calculation, we take the standard constituent quark masses,
namely $m_u$=$m_d$=330 MeV and $m_s$=450 MeV. The masses of the
final baryons are taken from PDG~\cite{Tanabashi:2018oca}, as listed
in Table.~\ref{paraments}. We adopt the simple harmonic oscillator
(SHO) wave functions for the space-wave functions of the hadrons.
The harmonic oscillator strength $\beta_{th}$ between the two
strange quarks for the initial mesons is determined by the spatial
wave functions obtained in mass spectrum calculations (as listed in
Table~\ref{MASS}). The harmonic oscillator strength between the two
light quarks for final baryons is taken as $\alpha=400$ MeV. As to
the strength of the quark pair creation from the vacuum, we adopt
the same value as in Ref.\cite{Godfrey:2015dva}, $\gamma=6.95$. The
uncertainty of the strength $\gamma$ is about 30\%
~\cite{Blundell:1996as,Godfrey:2015dia,Close:2005se,Li:2010vx}, and
the partial decay widths are proportional to $\gamma^4$. Thus our
predictions may bare a quite large uncertainty.

\subsection{$\Lambda\bar{\Lambda}$ decay mode}

\subsubsection{States around the $\Lambda\bar{\Lambda}$ threshold }\label{resultsb}

In 2007, the BABAR Collaboration measured the cross section for
$e^+e^-\rightarrow \Lambda\bar{\Lambda}$ from threshold up to 3
GeV~\cite{Aubert:2007uf} and observed a possible nonvanishing cross
section at threshold. Recently, the BESIII Collaboration published a
measurement of the process $e^+e^-\rightarrow
\Lambda\bar{\Lambda}$~\cite{Ablikim:2017pyl} with improved
precision. The Born cross section at $\sqrt{s}=2232.4$ MeV, which is
1.0 MeV above the $\Lambda\bar{\Lambda}$ mass threshold, is measured
to be $305\pm45^{+66}_{-36}$ pb, which indicates an obvious
threshold enhancement.

\begin{table}[h]
\caption{\label{width} The partial decay widths of the vector
strangonium with a mass of $M=2232$ MeV. }
\begin{tabular}{ccccccccccc}\hline\hline
\text{State}  ~~~~~~~~~&$\beta_{th}$(\text{MeV})~~~~~~~~~&$\Gamma_{\Lambda\bar{\Lambda}}$(\text{MeV})\\
$\psi(3^3S_1)$~~~~~~~~~&368~~~~~~~~~&5.84\\
$\psi(2^3D_1)$~~~~~~~~~&375~~~~~~~~~&$3.90\times10^{-6}$\\
\hline\hline
\end{tabular}
\end{table}


According to various model predictions (see Table~\ref{MASS}), there
are two strangeonium meson resonances $\phi(3^3S_1)$ and
$\phi(2^3D_1)$ with both masses around 2.2 GeV and $J^P=1^{--}$. As
a possible source of the observed threshold enhancement, it is
crucial to study the decay properties of the states $\phi(3^3S_1)$
and $\phi(2^3D_1)$.

We first explore the $\Lambda\bar{\Lambda}$ partial decay width of
the state $\phi(3^3S_1)$ and obtain
\begin{equation}
\Gamma[\phi(3^3S_1){\to}\Lambda\bar{\Lambda}] \sim 5.84\text{ MeV}
\end{equation}
with a mass of $M=2232$ MeV (see Table~\ref{width}). This partial
decay width is large enough to be observed in experiments, and
indicates that the observed threshold enhancement may arise from
this state. Although the phase space is suppressed seriously around
threshold, the transition amplitude for this decay mode is quite
large. Hence, the partial decay width of the $\Lambda\bar{\Lambda}$
mode for the state $\phi(3^3S_1)$ reaches several MeV. Considering
the uncertainties of the predicted mass, we study the variation of
the $\Lambda\bar{\Lambda}$ decay width as a function of the mass of
the state $\phi(3^3S_1)$. The decay width increases rapidly with the
mass in the range of (2233-2300) MeV.

Then, we investigate the decay properties of the state
$\phi(2^3D_1)$. Fixing the mass at $M=2232$ MeV, we get
\begin{equation}
\Gamma[\phi(2^3D_1){\to}\Lambda\bar{\Lambda}] \sim
3.90\times10^{-6}\text{ MeV}.
\end{equation}
This width seems too small to be observed in experiments. Combining
the predicted partial decay width of $\phi(3^3S_1)$, we further
obtain
\begin{equation}
\frac{\Gamma[\phi(3^3S_1){\to}\Lambda\bar{\Lambda}]}{\Gamma[\phi(2^3D_1){\to}\Lambda\bar{\Lambda}]}\sim
1.50\times10^{6}.
\end{equation}
The decay ratio of $\phi(3^3S_1)$ into the $\Lambda\bar{\Lambda}$
channel is about $\mathcal{O}(10^6)$ larger than that of
$\phi(2^3D_1)$ into the $\Lambda\bar{\Lambda}$ channel. Combining
their electronic decay width we calculated in Sec.~III, we obtain
that if the threshold enhancement reported by the BESIII
Collaboration in the process $e^+e^-\rightarrow
\Lambda\bar{\Lambda}$ were related to an unobserved strangeonium
meson resonance, this state should most likely be $\phi(3^3S_1)$.

Besides the uncertainties coming from the predicted mass and
harmonic oscillator strength $\beta_{th}$, the results of
$\phi(3^3S_1)$ and $\phi(2^3D_1)$ may have large uncertainties due
to their lower masses. At the hadron level, the energy of the
intermediate states with the spin parity $J^{PC}=1^{--}$, such as
molecular states $KK_1(1270)$, $K*(892)K^*_0(700)$,
$K*(892)K_1(1270)$, and $\phi(1020)a_0(980)$ and so on, is about 1.7
Gev$\sim$2.1 GeV. Thus the $E_k-E_A$ are small and sensitive to the
masses of the intermediates state. In this case, taking
$E_k-E_A$=2$m_q$ as a constant will introduce a large uncertainty in
this calculation.

\begin{table}[ht]
\caption{\label{decay2} The predicted partial decay widths of the
higher $\phi$ mesons with $J^{PC}=1^{--}$. The unit is MeV.  }
\begin{tabular}{ccccccccccccccc}\hline\hline
\text{State}~~~~~~~~&\text{Mass}~~~~~~~~&$\beta_{th}$~~~~~~~~&$\Gamma_{\Lambda\bar{\Lambda}}$\\
\hline
$\phi(4^3S_1)$~~~~~~~~&2627~~~~~~~~&351~~~~~~~~&1.81\\
$\phi(5^3S_1)$~~~~~~~~&2996~~~~~~~~&341~~~~~~~~&0.08\\
$\phi(6^3S_1)$~~~~~~~~&3327~~~~~~~~&334~~~~~~~~&0.84\\
$\phi(3^3D_1)$~~~~~~~~&2732~~~~~~~~&355~~~~~~~~&3.40\\
$\phi(4^3D_1)$~~~~~~~~&3079~~~~~~~~&344~~~~~~~~&0.27\\
$\phi(5^3D_1)$~~~~~~~~&3395~~~~~~~~&336~~~~~~~~&0.10\\
\hline\hline
\end{tabular}
\end{table}

\begin{figure*}[hpbt]
\centering \epsfxsize=11 cm \epsfbox{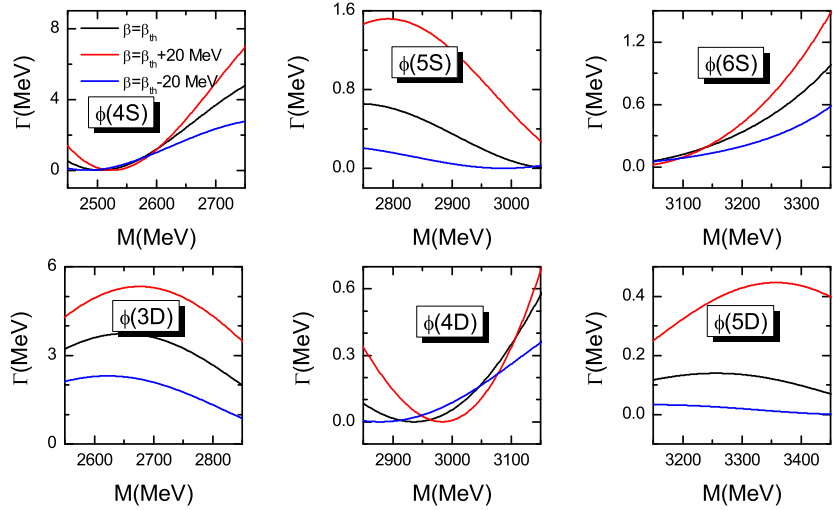} \caption{The
variation of the $\Lambda\bar{\Lambda}$ decay width with the mass of
the S- and D-wave vector strangeonium. The red, black, and blue
lines correspond to the predictions with different values of
$\beta$=$\beta_{th}+20$ MeV, $\beta_{th}$, and $\beta_{th}-20$ MeV,
respectively. }\label{SDWAVE}
\end{figure*}

\subsubsection{higher states}\label{resultsb}

Besides $\phi(3^3S_1)$ and $\phi(2^3D_1)$, we also analyze the decay
properties of the S-wave states $\phi(4^3S_1,~5^3S_1,~6^3S_1)$ and
the D-wave states $\phi(3^3D_1,~4^3D_1,~5^3D_1)$. The decay
properties are collected in Table~\ref{decay2}. From the table, we
get that the $\Lambda\bar{\Lambda}$ partial decay width of
$\phi(3^3D_1)$ can reach up to $\Gamma\sim3.5$ MeV, which is the
largest compared to five other vector states we considered in this
work. The sizeable width indicates that this state has a good
potential to be observed in the $\Lambda\bar{\Lambda}$ decay
channel.

Similarly, taking the uncertainties of the theoretical masses and
harmonic oscillator strength $\beta_{th}$ into account, we plot the
$\Lambda\bar{\Lambda}$ partial decay widths of those states as
functions of the masses in Fig.~\ref{SDWAVE} with different values
of $\beta$=$\beta_{th}$+20 MeV, $\beta_{th}$, and $\beta_{th}$-20
MeV, respectively. According to Fig.~\ref{SDWAVE}, for the state
$\phi(3^3D_1)$, the variation curve likes a bowel structure when the
mass varies from 2550 MeV to 2850 MeV, and the partial width can
reach up to $\Gamma\sim3.7$ MeV with $\beta=\beta_{th}$. The
$\Lambda\bar{\Lambda}$ partial decay width for $\phi(5^3D_1)$ is the
smallest. The decay width is less than $\Gamma<0.4$ MeV with the
mass in the range of $M=(3150-3450)$ MeV. As to $\phi(4^3S_1)$, its
$\Lambda\bar{\Lambda}$ decay width is very sensitive to the mass
(see Fig.~\ref{SDWAVE}). When $\beta=\beta_{th}$, the width varies
in the range of $\Gamma\sim(0.0-4.8)$ MeV with the mass in the range
of $M=(2450-2750)$ MeV. If the mass of $\phi(4^3S_1)$ lies in
(2496-2590) MeV, the decay width of the $\Lambda\bar{\Lambda}$ mode
is less than one MeV. Most of the $\Lambda\bar{\Lambda}$ partial
decay widths for the other three states, $\phi(4^3D_1)$,
$\phi(5^3S_1)$ and $\phi(6^3S_1)$, are less than one MeV (see
Fig.~\ref{SDWAVE}). These partial widths seem to be sizeable as
well.

\begin{table*}[htpb]
\caption{\label{other} The partial decay widths of the higher $\phi$
mesons with $J^{PC}=1^{--}$. The unit is MeV.}
\begin{tabular}{ccccccccccccccc}\hline\hline
\text{State}~~~~~~~~&\text{Mass}~~~~~~~~&$\beta$
~~~~~~~~&$\Gamma[\Xi^-\bar{\Xi}^+]$~~~~~~~~&$\Gamma[\Sigma^+\bar{\Sigma}^-]$~~~~~~~~&$\Gamma[\Sigma^{+*}\bar{\Sigma}^-]$
~~~~~~~~&$\Gamma[\Sigma^{+*}\bar{\Sigma}^{-*}]$~~~~~~~~&$\Gamma[\Xi^{-*}\bar{\Xi}^+]$\\
\hline
$\phi(4^3S_1)$~~~~~~~~&2627~~~~~~~~&351~~~~~~~~&$\cdot\cdot\cdot$~~~~~~~~&2.86~~~~~~~~&0.91~~~~~~~~&$\cdot\cdot\cdot$~~~~~~~~&$\cdot\cdot\cdot$\\
$\phi(5^3S_1)$~~~~~~~~&2996~~~~~~~~&341~~~~~~~~&1.28~~~~~~~~&0.05~~~~~~~~&0.62~~~~~~~~&1.69~~~~~~~~&0.03\\
$\phi(6^3S_1)$~~~~~~~~&3327~~~~~~~~&334~~~~~~~~&$2.26\times 10^{-3}$~~~~~~~~&0.08~~~~~~~~&0.06~~~~~~~~&$5.67\times 10^{-3}$~~~~~~~~&$7.63\times 10^{-4}$\\
$\phi(3^3D_1)$~~~~~~~~&2732~~~~~~~~&355~~~~~~~~&0.16~~~~~~~~&1.49~~~~~~~~&0.41~~~~~~~~&$\cdot\cdot\cdot$~~~~~~~~&$\cdot\cdot\cdot$\\
$\phi(4^3D_1)$~~~~~~~~&3079~~~~~~~~&344~~~~~~~~&0.11~~~~~~~~&0.02~~~~~~~~&0.07~~~~~~~~&0.86~~~~~~~~&0.01\\
$\phi(5^3D_1)$~~~~~~~~&3395~~~~~~~~&336~~~~~~~~&0.02~~~~~~~~&0.08~~~~~~~~&0.03~~~~~~~~&0.58~~~~~~~~&$9.55\times 10^{-5}$\\
\hline\hline
\end{tabular}
\end{table*}

\begin{figure*}[hpbt]
\centering \epsfxsize=15 cm \epsfbox{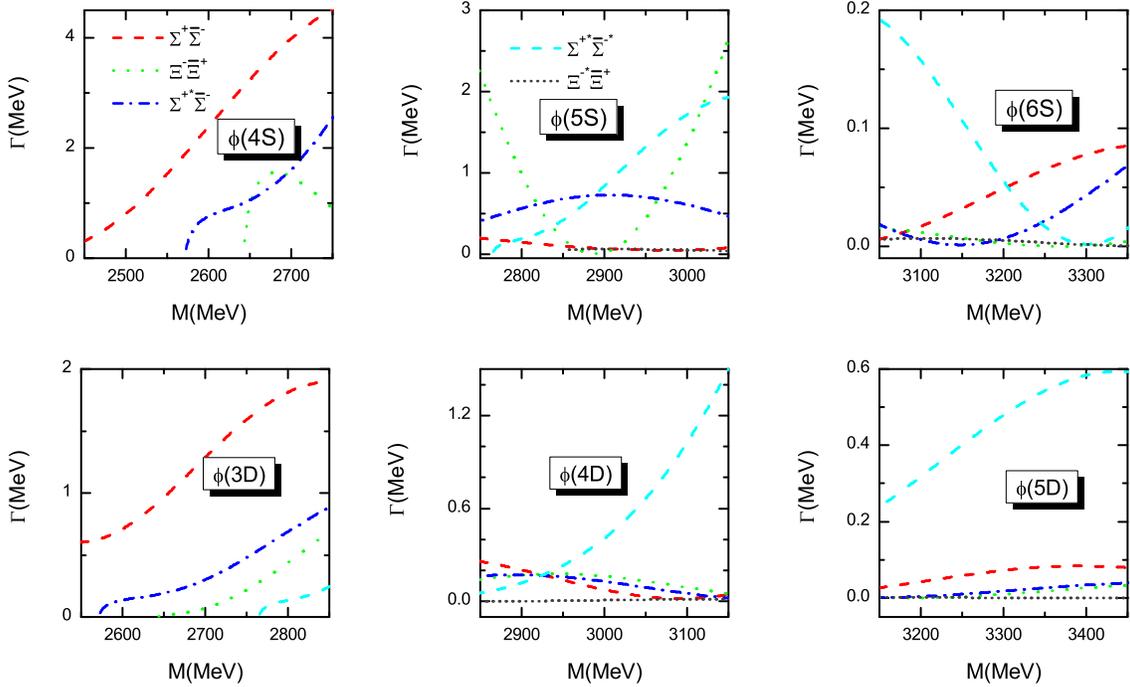} \caption{The
variation of the partial decay widths with the mass of the S- and
D-wave vector strangeonium. }\label{SDWAVEother}
\end{figure*}

\subsection{Other double baryon decay modes}
Besides $\Lambda\bar{\Lambda}$ decay mode, we also investigate the
$\Xi^{-(*)}\bar{\Xi}^+$ and $\Sigma^{+(*)}\bar{\Sigma}^{-(*)}$ decay
modes for the excited vector strangeonium. According to the
predicted masses listed in Table~\ref{MASS}, the masses of the
states $\phi(3S)$ and $\phi(2D)$ are under the threshold of
$\Xi^{-(*)}\bar{\Xi}^+$ and $\Sigma^{+(*)}\bar{\Sigma}^{-(*)}$.
Thus, in this section, we focus on partial decay properties of the
vector strangeonium states $\phi(4S,~5S,~6S)$ and
$\phi(3D,~4D,~5D)$. Our predictions are collected in
Table~\ref{other}.

From the Table, we notice that the $\Sigma^{+}\bar{\Sigma}^{-}$
partial decay width of $\phi(4^3S_1)$ and $\phi(3^3D_1)$ can reach
up to $\Gamma\sim2.9$ MeV and $\Gamma\sim1.5$ MeV, respectively,
which are large enough to be observed in future experiments.
Meanwhile, the $\Xi^{-}\bar{\Xi}^+$ and
$\Sigma^{+*}\bar{\Sigma}^{-*}$ partial decay widths of the state
$\phi(5^3S_1)$ are both larger than one MeV.

In addition, we also plot the decay properties of the states
$\phi(4S,~5S,~6S)$ and $\phi(3D,~4D,~5D)$ as a function of the mass
in Fig.~\ref{SDWAVEother}.

To investigate the uncertainties of the parameter $\beta_{th}$, we
further consider the partial decay properties with different
$\beta_{th}$ values. The theoretical numerical results are not shown
in the present work. According to our calculations, our main
predictions hold in a reasonable range of the parameter
$\beta_{th}$.

\section{Summary}\label{suma}

In the present work, we have studied the mass spectrum of the
strangeonium system with the GI model and further investigated the
electronic decay width and $\Lambda\bar{\Lambda}$,
$\Xi^{-(*)}\bar{\Xi}^+$, and $\Sigma^{+(*)}\bar{\Sigma}^{-(*)}$
double baryons decay widths of the excited vector strangeonium
states $\phi(3S,~4S,~5S,~6S)$ and $\phi(2D,~3D,~4D,~5D)$.

For the electronic decay widths, we obtain that the electronic decay
widths of the excited vector strangeonium states
$\phi(3S,~4S,~5S,~6S)$ and $\phi(2D,~3D,~4D,~5D)$ are smaller than
that of the state $\phi(1S)$. Meanwhile, the electronic decay width
of the $D$-wave vector strangeonium is about $3\sim8$ times larger
than that of the $S$-wave vector strangeonium.

For the double baryons decay widths, the partial decay width of the
$\Lambda\bar{\Lambda}$ mode can reach up to $\sim5.84$ MeV for
$\phi(3S)$, while the partial $\Lambda\bar{\Lambda}$ decay width of
the states $\phi(2D)$ is about $\mathcal{O}(10^{-3})$ keV. Thus, the
$\Lambda\bar{\Lambda}$ decay width ratio between the states
$\phi(3^3S_1)$ and $\phi(2^3D_1)$ is $\mathcal{O}(10^6)$. If the
threshold enhancement reported by the BESIII Collaboration in
process $e^+e^-\rightarrow \Lambda\bar{\Lambda}$ does arise from an
unobserved strangeonium meson, the resonance is most likely to be
the strangeonium state $\phi(3S)$. We also notice that the
$\Lambda\bar{\Lambda}$ and $\Sigma^{+}\bar{\Sigma}^{-}$ partial
decay widths of the states $\phi(3^3D_1)$ and $\phi(4^3S_1)$ are
about several MeV, respectively, which are enough to be observed in
future experiments. The double baryons decay modes provide a unique
probe of the excited vector strangeonium resonances, which may be
produced and investigated at BESIII and BelleII.

\section*{Acknowledgements }

We would like to thank Xiao-Lin Chen and Wei-Zhen Deng for very
helpful suggestions. We also thank Guang-Juan Wang and Lu Meng for
very useful discussions. This work is supported by the National
Natural Science Foundation of China under Grants No. ~11575008,
~11621131001, ~11775078, No.~U1832173 and National Key Basic
Research Program of China (2015CB856700). This work is also in part
supported by China Postdoctoral Science Foundation under Grant
No.~2017M620492.


\end{document}